\newcommand{\CLASSINPUTtoptextmargin}{50pt}
\newcommand{\CLASSINPUTbottomtextmargin}{50pt}
\newcommand{\CLASSINPUTinnersidemargin}{41pt}
\newcommand{\CLASSINPUToutersidemargin}{41pt}
\documentclass[9pt,sigconf,screen=true,bookmarks=false]{acmart}

\usepackage{color,xcolor}
\usepackage{xspace}
\usepackage{multirow}
\usepackage{booktabs}
\usepackage{lipsum}
\usepackage{soulutf8}
\usepackage{diagbox}
\usepackage{physics}

\usepackage{colortbl}

\usepackage{enumitem}
\usepackage{braket}
\usepackage{calc}

\usepackage{amsmath}
\usepackage{amsfonts}
\usepackage{url}

\usepackage{afterpage}
\usepackage{cancel}

\usepackage[ruled,vlined,noend]{algorithm2e}
\usepackage{setspace}

\usepackage{pifont}

\usepackage{titlesec}

\usepackage{hyperref}
\hypersetup{
    colorlinks=true,
    linkcolor=magenta,
    filecolor=magenta,      
    urlcolor=blue,
}

\usepackage{xcolor}
\usepackage{soul}

\definecolor{citecolor}{RGB}{34,139,34}
\definecolor{mydarkblue}{rgb}{0,0.08,1}
\definecolor{mydarkgreen}{rgb}{0.02,0.6,0.02}
\definecolor{mydarkred}{rgb}{0.8,0.02,0.02}
\definecolor{mydarkorange}{rgb}{0.40,0.2,0.02}
\definecolor{mypurple}{RGB}{111,0,255}
\definecolor{myred}{rgb}{1.0,0.0,0.0}
\definecolor{mygold}{rgb}{0.75,0.6,0.12}
\definecolor{myblue}{rgb}{0,0.2,0.8}
\definecolor{mydarkgray}{rgb}{0.,0.2,0.2}

\definecolor{lightred}{RGB}{255,235,235}
\definecolor{lightgreen}{RGB}{235,255,235}
\definecolor{lightblue}{RGB}{235,235,255}
\definecolor{lightcyan}{RGB}{235,255,255}
\definecolor{lightmagenta}{RGB}{255,235,255}
\definecolor{lightyellow}{RGB}{255,255,235}

\definecolor{qxkcolor}{RGB}{215,235,255}
\definecolor{softmaxcolor}{RGB}{230,235,255}
\definecolor{probxvcolor}{RGB}{255,255,235}

\definecolor{topkcolor}{RGB}{255,235,235}
\definecolor{zecolor}{RGB}{255,255,235}
\definecolor{dynacolor}{RGB}{235,255,255}

\definecolor{reviewcolor}{RGB}{0,0,200}

\renewcommand\footnotemark{}
\newcommand{\name}{Transformer-QEC\xspace}

\newcounter{rlabelno}

\definecolor{color1}{rgb}{1, 1, 0.9}
\definecolor{color2}{rgb}{1, 0.9, 1}
\definecolor{color3}{rgb}{0.9, 1, 1}

\definecolor{color4}{rgb}{1, 0.9, 0.9}
\definecolor{color5}{rgb}{0.9, 0.9, 1}
\definecolor{color6}{rgb}{0.9, 1, 0.9}

\definecolor{color7}{rgb}{0.8, 0.9, 1}
\definecolor{color8}{rgb}{0.9, 1, 0.8}
\definecolor{color9}{rgb}{1, 0.8, 0.9}

\newcolumntype{a}{>{\columncolor{color1}}c}
\newcolumntype{b}{>{\columncolor{color2}}c}
\newcolumntype{d}{>{\columncolor{color3}}c}
\newcolumntype{e}{>{\columncolor{color4}}c}
\newcolumntype{f}{>{\columncolor{color5}}c}
\newcolumntype{g}{>{\columncolor{color6}}c}
\newcolumntype{h}{>{\columncolor{color7}}c}
\newcolumntype{i}{>{\columncolor{color8}}c}
\newcolumntype{j}{>{\columncolor{color9}}c}

\usepackage{geometry}

\begin{document}
\settopmatter{printacmref=false} % Removes citation information below abstract
% \bstctlcite{IEEEexample:BSTcontrol}
\renewcommand\footnotetextcopyrightpermission[1]{} % removes footnote with conference information in first column

\pagestyle{fancy}
\fancyhead[L]{}
\fancyhead[R]{}

\title{\name: Quantum Error Correction Code Decoding \\ with Transferable Transformers}

\author{
Hanrui Wang$^{1}$, Pengyu Liu$^{2}$, Kevin Shao$^{1}$, Dantong Li$^{3}$,\\ Jiaqi Gu$^{4}$,  David Z. Pan$^{5}$, Yongshan Ding$^{3}$, Song Han$^{1}$\\
\small{
$^{1}$Massachusetts Institute of Technology, $^{2}$Carnegie Mellon University, $^{3}$Yale University, $^{4}$Arizona State University, $^{5}$University of Texas at Austin}
}

\pagenumbering{arabic}
\pagestyle{plain}

\begin{abstract}

Quantum computing has the potential to solve problems that are intractable for classical systems, yet the high error rates in contemporary quantum devices often exceed tolerable limits for useful algorithm execution. Quantum Error Correction (QEC) mitigates this by employing redundancy, distributing quantum information across multiple data qubits and utilizing syndrome qubits to monitor their states for errors. The syndromes are subsequently interpreted by a decoding algorithm to identify and correct errors in the data qubits. This task is complex due to the multiplicity of error sources affecting both data and syndrome qubits as well as syndrome extraction operations. Additionally, identical syndromes can emanate from different error sources, necessitating a decoding algorithm that evaluates syndromes collectively. Although machine learning (ML) decoders such as multi-layer perceptrons (MLPs) and convolutional neural networks (CNNs) have been proposed, they often focus on \textit{local} syndrome regions and require \textit{retraining} when adjusting for different code distances.
To address these issues, we introduce a transformer-based QEC decoder, termed \mbox{\name}, which employs self-attention to achieve a global receptive field across all input syndromes. It incorporates a \textit{mixed loss} training approach, combining both local physical error and global parity label losses. Moreover, the transformer architecture's inherent adaptability to variable-length inputs allows for efficient \textit{transfer learning}, enabling the decoder to adapt to varying code distances without retraining.

Evaluation on six code distances and ten different error configurations demonstrates that our model consistently outperforms non-ML decoders, such as Union Find (UF) and Minimum Weight Perfect Matching (MWPM), and other ML decoders, thereby achieving best logical error rates. Moreover, the transfer learning can save over 10$\times$ of training cost. 

\end{abstract}

\maketitle

\section{Introduction}

\begin{figure}[t]
    \centering
    \includegraphics[width=\columnwidth]{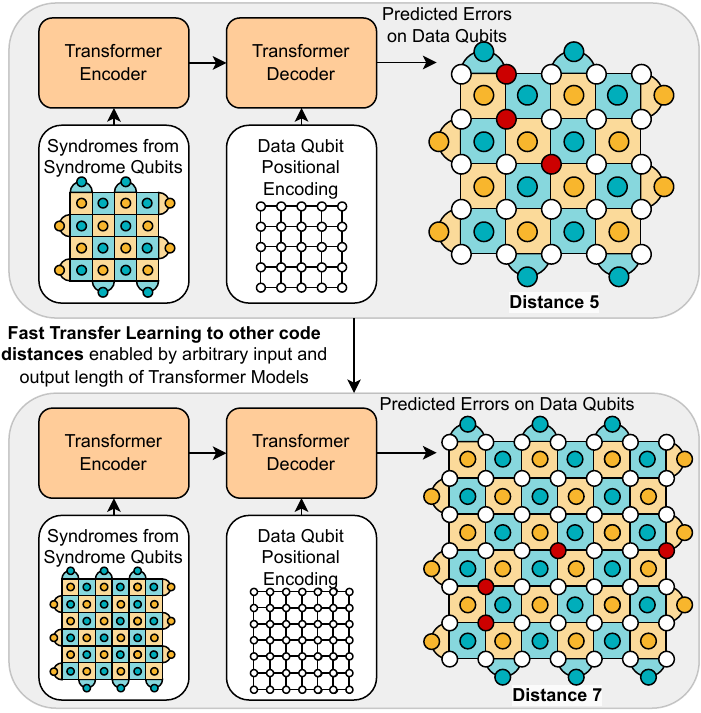}
    % \vspace{-15pt}
    \caption{Transformer for error correction decoding overview. The Transformer takes syndrome inputs and processes them through both the transformer model encoder and decoder. The output of this process consists of error predictions. One notable advantage is its ability to be seamlessly applied to different code distances due to the transformer model's flexible input and output size.}
    % \vspace{-20pt}
    \label{fig:teaser}
\end{figure}

The field of Quantum Computing (QC) has attracted significant attention in research circles as a novel computational paradigm, poised to tackle challenges previously beyond the reach of conventional computing with remarkable efficiency. QC offers promising prospects across various industries and academic fields, notably impacting areas such as cryptography~\cite{shor1999polynomial}, database searching~\cite{grover1996fast}, combinatorial optimization~\cite{farhi2014quantum, liang2023hybrid}, molecular dynamics simulations~\cite{peruzzo2014variational}, and advancements in machine learning~\cite{lloyd2013quantum, liang2021can, wang2022quantumnas, wang2022quantumnat, wang2022qoc, wang2022quest, wang2023robuststate, zheng2022sncqa, cheng2022topgen, liang2022pan}, among others.

Despite significant progress in quantum computing hardware~\cite{ibm127}, current devices exhibit high error rates, ranging from $10^{-3}$ to $10^{-2}$, which are orders of magnitude above the thresholds required for practical applications (below $10^{-10}$)~\cite{lee2021even}. Addressing these error rates is crucial for advancing the field. Quantum Error Correction (QEC) serves as a key technique for error mitigation by incorporating redundancy—distributing the information of a single logical qubit across multiple physical qubits. With adequate redundancy in QEC codes, the logical error rate can decrease exponentially, provided the physical error rate \( p \) remains below a specific threshold. Thus, judiciously managed redundancy via QEC can enable quantum systems to achieve the low error rates necessary for practical computation. In QEC, logical qubits encode across multiple data qubits, with error detection aided by parity qubits, as shown in Figure~\ref{fig:teaser}. Periodic syndrome extraction localizes errors into detectable syndromes over several cycles. A classical decoder then interprets these syndromes to correct errors, influenced by physical error rates and decoder performance. Decoders often require a broad syndrome field due to potential syndrome overlap from different errors.

\begin{figure*}[t]
    \centering
    \includegraphics[width=\textwidth]{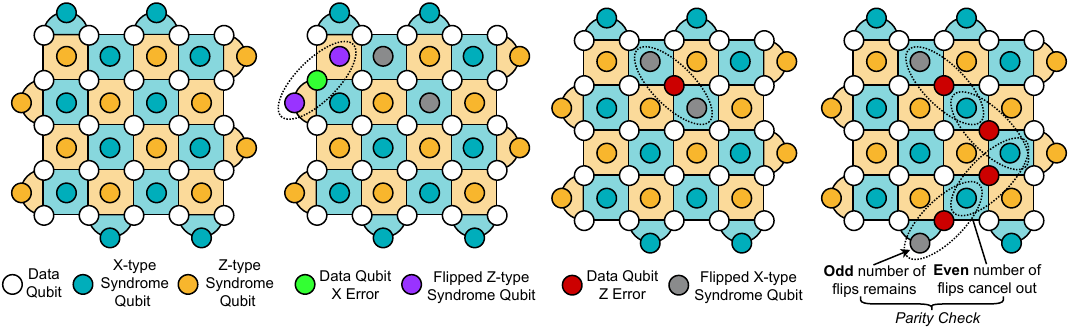}
    % \vspace{-15pt}
    \caption{The surface code contains data qubits and two kinds of syndrome qubits. X-type syndrome qubits in green checks Z errors while Z-type syndrome qubits in yellow check X errors. When error occurs on data qubits, the nearby syndromes may be flipped depending on the parity of data qubits. When multiple error occurs, the syndrome patterns will be more difficult to decode.}
    % \vspace{-20pt}
    \label{fig:surface}
\end{figure*}

The rotated surface code~\cite{horsman2012surface} is a promising candidate for realizing fault tolerance. Well-established algorithms for surface code include Minimum Weight Perfect Matching (MWPM) and Union Find (UF). Recently, machine learning (ML), especially neural network (NN) based decoders have gained attention due to a few desirable characteristics. First, they generally run in constant time, which is necessary to prevent a backlog of syndrome outcomes. Second, unlike MWPM, they are capable of learning both correlations between physical errors (such as the correlation between X and Z error in depolarizing errors) as well as learning hidden and potentially changing underlying physical error distributions.

However, ML based decoders also bring significant challenges. First, several models, such as those grounded in convolutional neural networks (CNN), are limited by a small receptive field. This constraint may hinder their ability to accurately pinpoint long error chains. Second, many models, like the multi-layer perceptron (MLP), has a fixed size for both input and output. As a result, changes in code distance would mandate retraining of an entirely new model, leading to considerable overhead.

Therefore, to solve these challenges, we propose \name, a transferable transformer model designed for accurate and efficient decoding of surface code, as illustrated in Fig.~\ref{fig:teaser}. For the sake of simplicity, the figure only depicts two dimensions, but in reality, the input syndromes include an additional temporal dimension -- $round$ as in Fig.~\ref{fig:surface}. Our proposed model employs a transformer structure, incorporating both an encoder and a decoder to process the syndromes. Binary features on each syndrome qubit are projected to token embeddings and augmented with a 3D sinusoidal positional encoding, informing the model about the location of each qubit. The embeddings of the 3D inputs are then flattened to 1D input sequence and processed by the transformer encoder layer. Thanks to the global interaction capability brought by attention layer, all input syndromes can be considered holistically which boosts accuracy. The decoder then uses the positional encoding of the data qubits to predict the X or Z errors on each of them. Moreover, we propose a \textit{mixed loss} that combines the loss from the local physical error of each qubit with the loss from global parity prediction.

To optimize computational efficiency in the context of varying code distances for quantum error correction, we advocate for a transfer learning paradigm. Taking advantage of the transformer model's capacity for arbitrary input lengths, we repurpose pretrained models for different code distances by merely modifying the input sequence. For example, a model trained initially for a code distance $d=5$ can be efficiently fine-tuned for alternate distances ($d=7$ or $d=9$), thereby achieving enhanced performance metrics. Our approach yields a tenfold reduction in computational costs as compared to training from scratch, as verified through our experiments.

\begin{figure}[t]
    \centering
    \includegraphics[width=\columnwidth]{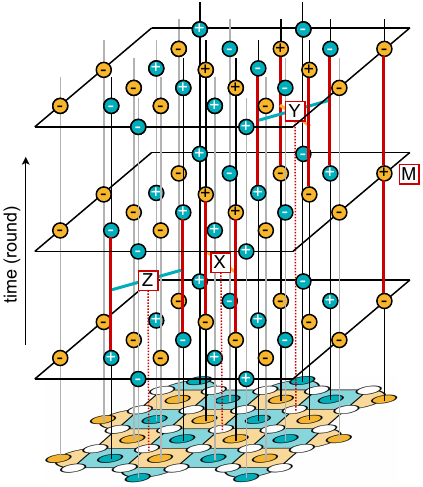}
    % \vspace{-15pt}
    \caption{Multiple rounds of surface code measurement. The progression of time is depicted by moving upwards from the array at the base, with each horizontal plane representing a step in the measurement process. In reality, errors will also occur in the syndrome extraction circuit and syndrome qubits, necessitating the need to repeat multiple rounds for decoding. On the right side, the measurement error on the syndrome qubit will also flip the syndrome.}
    % \vspace{-20pt}
    \label{fig:surface_3d}
\end{figure}

\begin{figure*}[t]
    \centering
    \includegraphics[width=\textwidth]{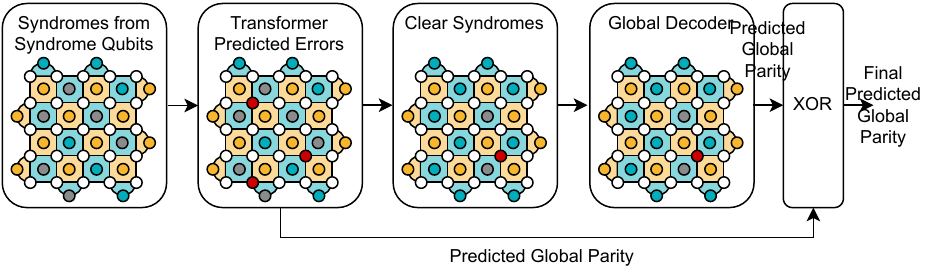}
    % \vspace{-15pt}
    \caption{Overall workflow of \name. The syndromes are firstly processed by the transformer model to predict the errors. Since the errors may not fully clear all syndromes, we will pass the cleared syndromes to a global decoder to predict a global parity. The final global parity is the XOR of the global parity from transformer predicted physical error and that predicted by global decoder.}
    % \vspace{-20pt}
    \label{fig:flow}
\end{figure*}

\begin{figure}[t]
    \centering
    \includegraphics[width=\columnwidth]{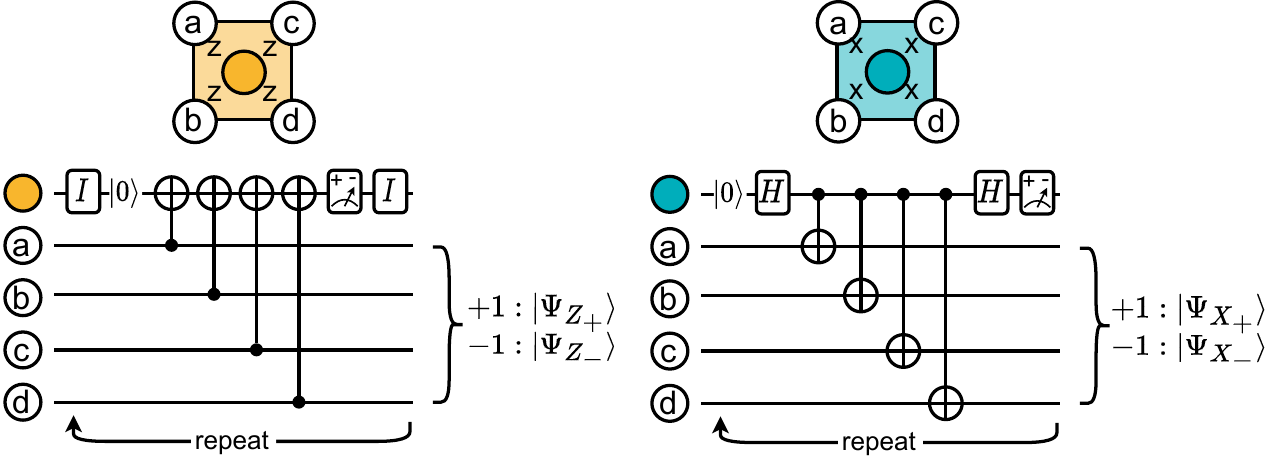}
    % \vspace{-15pt}
    \caption{Syndrome extraction circuit. Top: Z-type syndrome qubits. Bottom: X-type syndrome qubits.}
    % \vspace{-20pt}
    \label{fig:extraction}
\end{figure}

We extensively evaluate \name across six code distances, 3, 5, 7, 9, 11 and 13 and compare it with MWPM, UF and MLP baselines under 10 different error rates. Our results demonstrate that \name consistently surpasses these baselines, achieving the lowest logical error rates. In summary, \name makes four key contributions:
\begin{itemize}
    \item \textbf{A novel transformer-based model} for surface code decoding which uses the syndrome with positional encoding as inputs and predict errors.
    \item \textbf{A mixed loss} approach combined loss from local physical error prediction and global parity prediction improves the model's trainability and performance.
    \item \textbf{Transfer learning across different code distances}. For the first time, we propose to transfer the knowledge learn on one distance to another, thus reducing costs.
    \item \textbf{Extensive evaluations} on different physical error rates and distances demonstrates that our model consistently outperforms baselines such as MWPM, UF and MLP.
\end{itemize}

\section{Background and Related Works}

\noindent\textbf{Quantum error correction.}
Quantum Error Correction (QEC) enhances logical qubit fidelity by expanding into multiple physical qubits. The method involves syndrome extraction and error decoding steps. Using auxiliary syndrome qubits, errors are detected and classified into discrete Pauli categories~\cite{nielsen2002quantum}. If the physical error rate is below a certain threshold, QEC reduces the logical error rate, requiring more physical qubits. Quantum techniques often have added complexity compared to classical ones, with specific schemes viable for NISQ devices. This work focuses on the rotated surface code.

\noindent\textbf{Surface code.} The surface code is a leading QEC scheme that employs a two-dimensional lattice of interlaced data and syndrome qubits to encode a logical qubit. Notable for its high error threshold and need for only nearest-neighbor connectivity, it offers practicality for real-world quantum systems. The code distance $D$ dictates the lattice size and correlates with error resilience. Adjacent parity qubits detect errors on data qubits via a stabilizer circuit, capable of identifying X, Z, or Y errors, as depicted in Fig.~\ref{fig:extraction}. 
The code can correct error chains up to a length of $\lfloor \frac{D-1}{2} \rfloor$. A simplified variant, the 'rotated' surface code (Fig.~\ref{fig:surface}), reduces qubit and gate overhead and is often favored in practice. Characterized as a $[[D^2, 1, D]]$ stabilizer code, it stands as a viable option for near-term fault-tolerant quantum computation. Its design allows transversal single-qubit operations and enables two-qubit CNOT gates via lattice surgery~\cite{Horsman_2012}, thereby enhancing its feasibility for physical implementations. Decoders analyze syndromes from ancilla measurements to correct data qubit errors. They independently address X-type and Z-type errors, implicitly fixing Y-type errors. For large-scale FTQCs, decoders must be accurate, fast, and scalable~\cite{das2022afs}. Accuracy indicates reliable error detection, latency mandates cycle-limited operation, and scalability ensures efficient resource use. Increased accuracy may prolong operation time.

Decoders interpret syndromes, outcomes of ancilla measurements as shown in Fig.\ref{fig:surface_3d}, to identify and correct errors in data qubits. X-type and Z-type errors are treated independently, which inherently rectifies Y-type errors. Effective decoders for large-scale fault-tolerant quantum computers (FTQCs) must satisfy three primary criteria: accuracy, latency, and scalability\cite{das2022afs}. Accuracy denotes the decoder's reliability in error identification. Latency requires the decoder to function within a single syndrome extraction cycle. Scalability entails efficient resource utilization, crucial for hardware-constrained environments. Notably, increased accuracy often comes at the cost of extended operational time.

\noindent\textbf{ML based decoder.} The landscape of quantum computing has been significantly enriched by ML-based decoders, notably through Neural Network (NN) and Reinforcement Learning (RL) paradigms. In the realm of NNs, Boltzmann machines initiated ML-based decoding in toric codes~\cite{PhysRevLett.119.030501}, later broadened by Multi-Layer Perceptrons~\cite{Chamberland_2018} and Long Short-Term Memory networks for surface codes~\cite{Baireuther2018machinelearning}. On the RL front, advancements include transforming decoding into an RL environment~\cite{Sweke_2020} and optimizing RL decoders for specific error types~\cite{Andreasson2019quantumerror}. Efforts for scalable decoding have ranged from low-depth Convolutional Neural Networks~\cite{Breuckmann2018scalableneural} to multilevel architectures~\cite{Wagner_NN_Symmetries}. Guided by these contributions, our research focuses on employing Transformer-based NNs for local decoding, aiming to develop novel architectures and enhance scalability through a two-level local-global design. There exist numerous hardware accelerators~\cite{wang2020efficient, wang2021spatten, wang2023spattenchip, lin2021pointacc, wang2020sparch} that can be used to improve the efficiency and speed of NN-based decoders.

\begin{figure*}[t]
    \centering
    \includegraphics[width=\textwidth]{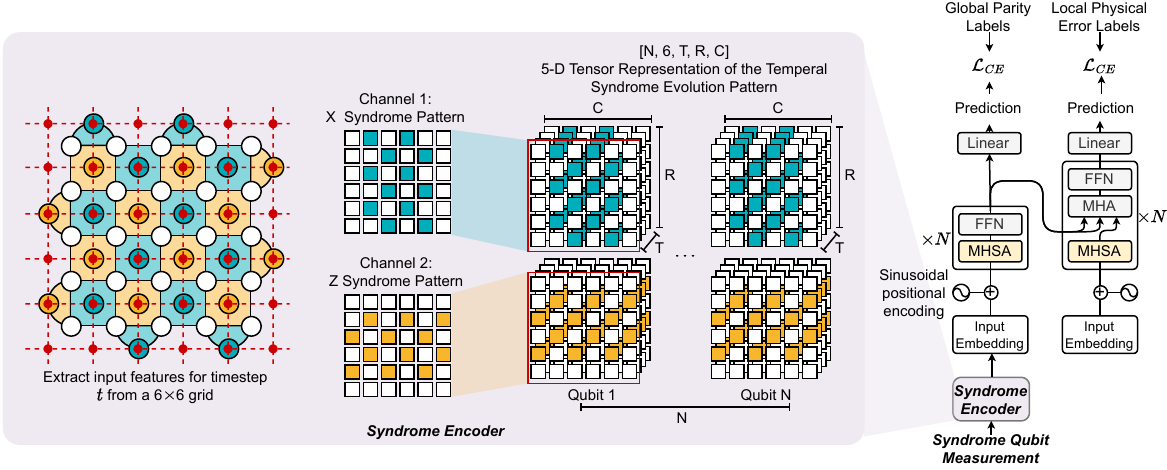}
    % \vspace{-15pt}
    \caption{Transformer model architecture. The input of the syndromes will be encoded by a ($D+1$) cubic grid. The input will go through the transformer encoder with self attention and FFN layers. Then the transformer decoder will produce the physical error predictions by processing the positional encoding of data qubits with size $D$ cubic.}
    % \vspace{-20pt}
    \label{fig:transformer}
\end{figure*}

\noindent\textbf{Non-ML based decoder.} Various decoding algorithms for QEC offer distinct advantages. Minimum Weight Perfect Matching (MWPM) uses Edmonds' method for optimal error syndrome pairing in topological codes~\cite{fowler2012surface, wu2023fusion}. The Union Find (UF) decoder features linear-time complexity for toric and surface codes~\cite{delfosse2017linear}. Lookup Table (LUT) decoders rely on predetermined error patterns~\cite{TomitaSvore}, while Tensor Network (TN) decoders utilize tensor structures for high error thresholds in topological codes~\cite{chubb2021general}. MWPM decoders~\cite{edmonds1965maximum, dennis2002topological, fowler2012surface, holmes2020nisq} are considered to have a good balance between the decoding accuracy and speed. It has almost linear time complexity~\cite{fowler2012towards, higgott2023sparse, wu2023fusion}, is practical for hardware implementation for real-time decoding~\cite{vittal2023astrea} and is more accurate than LUT and Union Find decoders. ~\cite{wang2023dgr} proposes to tackle the noise drift by updating the MWPM decoding graph weights.

\section{Methodology}

In this section, we delineate the error correction pipeline of our \name framework, followed by detailed discussions on the incorporated transformer model and transfer learning.

\noindent\textbf{Overall workflow.}
The iterative syndrome extraction procedure yields syndrome outcomes at discrete rounds, serving as the basis for the decoder's error predictions. Purely ML-based decoders, while effective, may yield predictions misaligned with the original syndromes. Therefore, an auxiliary, non-ML decoder is employed for syndrome clearance, as depicted in Figure~\ref{fig:flow}. Post-ML decoding, the predicted errors are utilized to reconcile syndromes and ascertain their global parity. Any residual errors are addressed by a global decoder like MWPM, ensuring complete syndrome clearance and yielding an additional global parity. The decoding process culminates in the XOR of the two global parities obtained.

\noindent\textbf{Transformer model.}
Given that the computational complexity of global decoders such as MWPM is often proportional to the number of non-zero syndromes, the efficiency of a preceding ML-based decoder in clearing syndromes can be highly advantageous. To this end, we introduce a novel Transformer-based decoder, schematically represented in Figure~\ref{fig:transformer}.
To encode input syndromes, we employ a cubic grid framework. For a surface code characterized by a distance $D$, a $D + 1$ square grid is utilized to ensure that each syndrome qubit resides at a grid intersection. Typically, the number of iterative rounds for syndrome extraction is set equal to the code distance. In our architecture, we incorporate an additional layer dedicated to the final syndrome measurement, thereby extending the round dimension to $D + 1$. Consequently, the feature space manifests as a $D + 1$ cubic grid. Each cell within this grid encapsulates a six-dimensional feature vector. The initial two dimensions specifically indicate the positions of the X-check and Z-check syndrome qubits, as further illustrated in Fig.~\ref{fig:transformer}.

Location encoding for the X and Z check syndrome qubits:
$$
\begin{bmatrix}
0 & 1 & 0 & 1 & 0 & 0 \\
0 & 0 & 1 & 0 & 1 & 0 \\
0 & 1 & 0 & 1 & 0 & 0 \\
0 & 0 & 1 & 0 & 1 & 0 \\
0 & 1 & 0 & 1 & 0 & 0 \\
0 & 0 & 1 & 0 & 1 & 0 \\
\end{bmatrix}
\begin{bmatrix}
0 & 0 & 0 & 0 & 0 & 0  \\
0 & 1 & 0 & 1 & 0 & 1  \\
1 & 0 & 1 & 0 & 1 & 0  \\
0 & 1 & 0 & 1 & 0 & 1  \\
1 & 0 & 1 & 0 & 1 & 0  \\
0 & 1 & 0 & 1 & 0 & 0  \\
\end{bmatrix}
$$
Subsequent to the positional dimensions, the next two channels in the six-dimensional feature vector encode the syndromes, varying across iterative rounds. For ultimate error correction, data qubits are measured in a designated basis. To enable network generalization across different syndrome measurement rounds, we define temporal lattice boundaries. Specifically, the fifth channel is initialized to 1 for the inaugural round and 0 for later rounds. Analogously, the sixth channel is set to 1 solely for the terminal round and 0 otherwise.

The feature vectors undergo dimensionality elevation via a learnable embedding layer. Subsequently, 3D sinusoidal positional encodings are integrated to convey qubit locations. This enhanced representation is then reshaped from 3D to 1D before forwarding to the Transformer encoder. The encoder comprises multiple layers, each featuring one multi-head self-attention (MHSA) module and one feed-forward network (FFN). The MHSA facilitates long-range contextual awareness by allowing each syndrome to attend to any other within the 3D grid. The FFN consists of two fully-connected layers, serving to further elevate the feature dimensionality, apply an activation function, and project back.

To predict physical errors, we employ Transformer decoder layers, using the positional encodings of data qubit locations as inputs. Each decoder layer comprises a self-attention module and a cross-attention module interfacing with the encoder. Queries for cross-attention originate from the decoder inputs, while keys and values are sourced from the encoder, thereby enabling the decoder to incorporate preceding syndrome information. Subsequently, a FFN layer and a prediction layer output the logits. Owing to the higher cost of false positives relative to false negatives, a confidence threshold is used for error prediction. Specifically, post-Sigmoid confidence $>0.95$ triggers a positive prediction.

\noindent\textbf{Mixed loss:} We introduce a composite loss function during training, integrating contributions from two distinct aspects. The first component arises from the prediction of local physical errors, while the second pertains to global parity prediction. The latter is computed via global average pooling of encoder output embeddings, subsequently processed through a prediction layer, as depicted in Figure~\ref{fig:transformer} (top right). This global parity serves as auxiliary information, enhancing the model's generalization across diverse syndrome patterns.

\noindent\textbf{Transfer learning:} Each code type encompasses a family of codes with varying distances, which necessitates different logical error rates based on the quantum algorithm deployed. Traditional approaches typically require retraining for each new code distance, incurring substantial computational and temporal overhead. To mitigate this, \name employs a transfer learning strategy, capitalizing on the inherent similarities between codes of different distances. For instance, handling syndromes in a distance-5 code bears resemblance to managing a sub-block in a distance-7 code. Utilizing a pre-trained model, we fine-tune it on the new distance's dataset. This flexibility is enabled by the Transformer architecture's inherent ability to accommodate sequences of arbitrary lengths. The sole aspect warranting careful adjustment is the positional encoding for the new distance.

\section{Evaluation}
\begin{table}[t]
\centering

\renewcommand*{\arraystretch}{1}
\setlength{\tabcolsep}{6pt}

\caption{Comparison of logical error rates under different code distance and physical error rates.}
\label{tab:main_results}
\resizebox{0.48\textwidth}{!}{%
\begin{tabular}{cc|cccc}
\toprule
 \multicolumn{2}{c|}{} & \multicolumn{4}{c}{Logical Error Rate $\downarrow$}\\
Distance & Phys. Err. Rate & UF & MWPM & MLP & \textbf{\name} \\
\midrule
 \multirow{2}{*}{3}& 0.0500               & 0.16745              & 0.14063              & 0.14794              & \textbf{0.13005}                       \\
 & 0.0100               & 0.01039              & 0.00800              & 0.00903              & \textbf{0.00784}              \\
 \midrule
 \multirow{2}{*}{5}& 0.0500               & 0.24120              & 0.17279              & 0.20888              & \textbf{0.17232}              \\
 & 0.0100               & 0.00406              & 0.00268              & 0.00443              & \textbf{0.00254}              \\
 \midrule
 \multirow{2}{*}{7}& 0.0500               & 0.29813              & 0.20178              & 0.28454              & \textbf{0.20590}              \\
 & 0.0100               & 0.00113              & 0.00064              & 0.00197              & \textbf{0.00059}              \\
 \midrule
 \multirow{2}{*}{9}& 0.0500               & 0.35250              & 0.23161              & 0.32770              & \textbf{0.23144}              \\
 & 0.0100               & 0.00028              & 0.00002     & 0.00017              & \textbf{0.00001}              \\
  
\bottomrule
\end{tabular}%
}
\end{table}

\subsection{Evaluation Methodology}

\noindent\textbf{Benchmarks.} We have selected the rotated surface code with distances of 3, 5, 7, 9. The round is set to be the same as the distance. The phenomenological error model~\cite{dennis02} we use encompasses errors on syndrome measurement and data qubits. Each syndrome qubit experiences a measurement error with a probability $p$. The errors on data qubits are depolarizing errors, which occur with a probability $p$, causing Pauli X, Y, or Z errors with equal probability. As assumed in previous work~\cite{das2021lilliput}, the error probabilities of these two types are considered to be equal. We choose values of $p$ from the set {0.05, 0.04, 0.03, 0.025, 0.02, 0.015, 0.01, 0.0075, 0.005, 0.0025}. The Google Stim package is used to construct the circuit and perform stabilizer simulations.

\noindent\textbf{Baselines.} Our three baselines include the Union Find decoder, the MWPM decoder as implemented in~\cite{wu2023fusion}, and a MLP architecture. Following \cite{Overwater_2022}, our MLP architecture has two hidden layers, with the dimensions of these layers set empirically. As the MLP requires fixed-size inputs and generates fixed-size outputs, it does not facilitate transfer learning like the Transformer does.

\noindent\textbf{Training settings.} Our main model is a Transformer with 6 layers, an embedding dimension of 256, 8 heads, and a FFN hidden dimension of 512. This model contains 7.9 million parameters. We also have a smaller model with 6 layers, an embedding dimension of 64, 2 heads, and an FFN hidden dimension of 128, which includes 0.5 million parameters. For training, we collect a dataset of 1,000,000 samples with a 1\% error rate. We use a learning rate of 0.001 with linear warmup and cosine decay, a weight decay with lambda 0.0001, and we train for 100 epochs. We utilize the Adam optimizer with a weighted binary cross-entropy loss for local physical errors, and a normal binary cross-entropy loss for global parity errors. For the MLP model, we use a physical error rate of 1\% for $d=3, 5$ and 2.5\% for $d=7, 9$. Like the initial Transformer model, we train for 100 epochs with the Adam optimizer. Training is conducted on a single NVIDIA A6000 GPU.
\begin{figure}[t]
    \centering
    \includegraphics[width=\columnwidth]{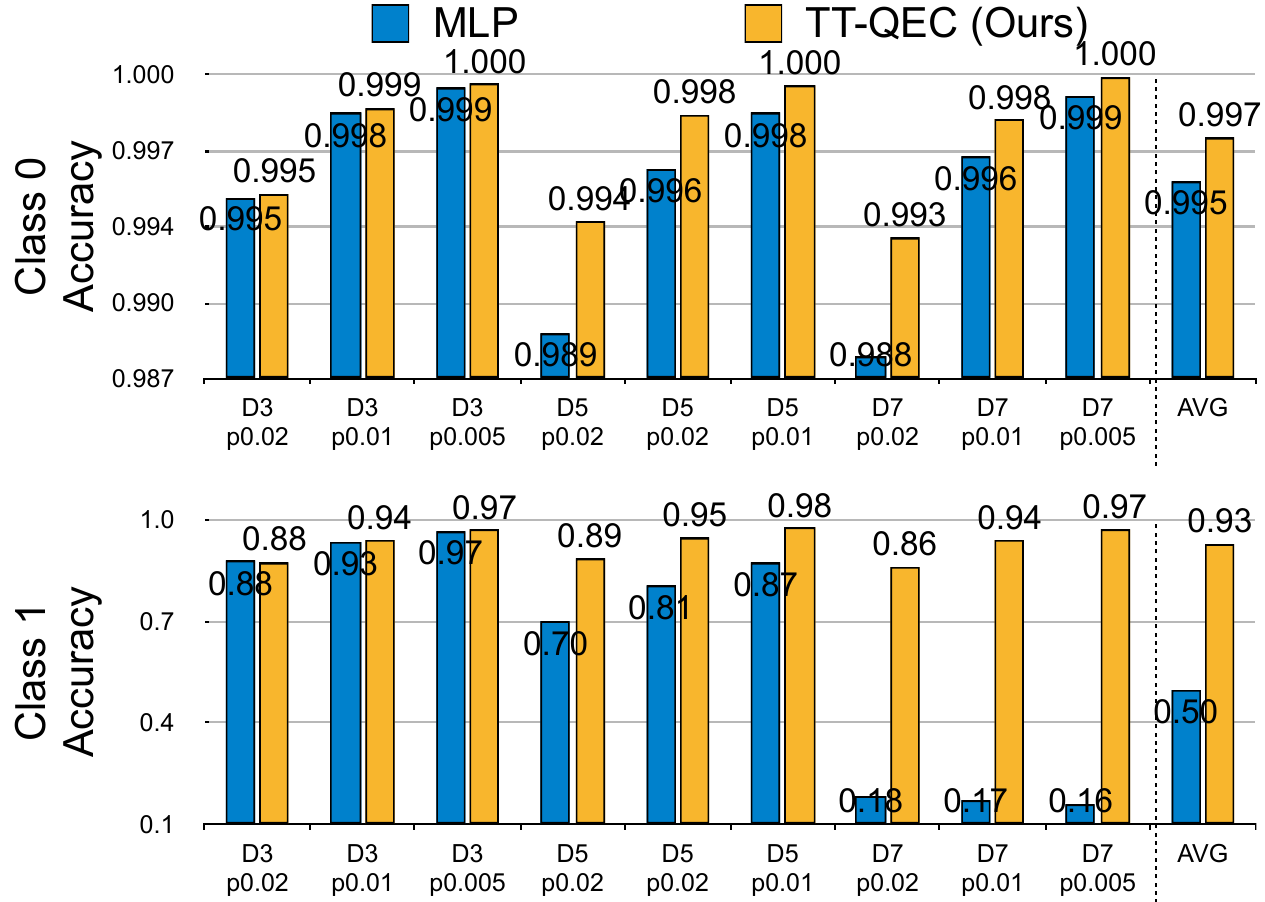}
    \vspace{-15pt}
    \caption{Accuracy comparison between the \name an MLP baseline. Class 0 accuracy is the accuracy of correctly identify a no error data qubit as no error (True Negative). Class 1 accuracy is the accuracy of correctly identify an error when the data qubit has error (True Positive).}
    \vspace{-10pt}
    \label{fig:vs_mlp}
\end{figure}

\noindent\textbf{Transfer learning settings.} The distance 5 model, trained from scratch, is used as the source model for transfer learning. For all other distances, we use a constant learning rate of 0.0005 and train for 10 epochs. All other settings remain identical to the settings of training from scratch.

\subsection{Experiment Results}

\begin{figure}[t]
    \centering
    \includegraphics[width=\columnwidth]{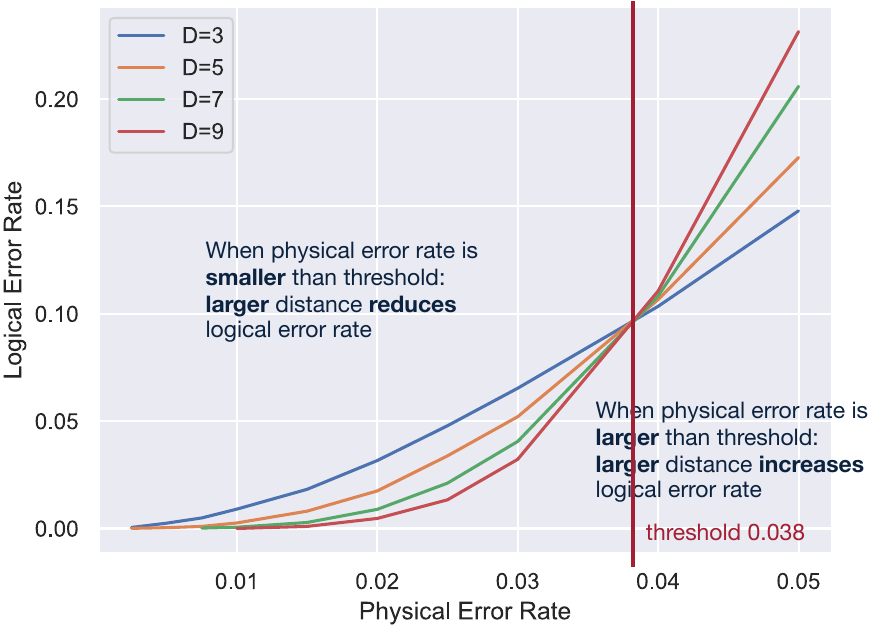}
    % \vspace{-15pt}
    \caption{Threshold of transformer based decoder. The threshold indicated the largest acceptable physical error rate for which using QEC can reduce error rate. Transformer obtains about 0.038 threshold.}
    \vspace{10pt}
    \label{fig:threshold}
\end{figure}

\noindent\textbf{Main results.} Table~\ref{tab:main_results} presents our primary results for varying code distances and physical error rates. Notice that all the models are transferred from the distance 5 model. In general, \name achieves a lower logical error rate for all benchmarks. The improvements over Union Find and MLP decoders are considerably more significant than the MWPM. This is likely because the global decoder in \name's framework also employs an MWPM. The MLP model can surpass the Union Find but is generally not as good as MWPM and \name, even though the MLP models are trained individually for each code distance. This result highlights the effectiveness of our proposed transfer learning techniques. Furthermore, in Figure~\ref{fig:vs_mlp}, we show the physical error prediction accuracy for baseline MLP and our \name. The class 0 accuracy means the ratio of predicted 0 when the ground truth is 0 (true negative). The class 1 accuracy means the ratio of predicted 1 when the ground truth is 1 (true positive). We can see that the accuracy for class 0 is in general much higher then class 1 because of the imbalance of training dataset. Moreover, our \name can achieve 43\% higher accuracy on the class 1 which means the \name model can identify errors with much higher reliability. That is beneficial when we desire the low level decoder to clear as many as syndromes as possible and speedup the end-to-end process.

\noindent\textbf{Evaluation of the threshold.} Figure~\ref{fig:threshold} shows the threshold evaluation of \name, with the X-axis as physical and Y-axis as logical error rates. The curves of different distances intersect at a point where the physical error rate is 0.0038 and the logical error rate is around 0.09. When $p$ is smaller than the threshold, larger code distances reduce the logical error rate. However, when $p$ is larger than the threshold, larger distances do not help. Instead, we observe larger logical error rates. This trend can be attributed to the increased error introduced by larger system sizes, which eclipses the benefits of greater redundancy with more qubits.

\noindent\textbf{Effectiveness of mixed loss.} To evaluate the mixed loss function, we perform an ablation study on the distance 5 code, as shown in Table~\ref{tab:two_branch}. Each column shows the comparison of the logical error rate under a specific physical error rate. The performance with both local and global loss can achieve better or equivalent performance for nine out of ten cases. This demonstrates that the global parity loss provides valuable guidance during the model's training process.

\noindent\textbf{Ablation on model size.} We evaluate two models with different sizes but the same training setting in Table~\ref{tab:model_size} under code distance 5 and varying physical error rates. It is evident that the larger model, with approximately 8 million parameters, outperforms the smaller model with 500 thousand parameters. The larger model is not overfitted to the training set and performs poorly on testing. This outcome is mainly due to the large size of the training set.

\begin{table}[t]
\centering

\renewcommand*{\arraystretch}{1}
\setlength{\tabcolsep}{6pt}

\caption{Comparison of logical error rate with global loss.}
% \vspace{-10pt}
\label{tab:two_branch}
\resizebox{\columnwidth}{!}{%
\begin{tabular}{lccccc}
\toprule
Error Rate & 0.0500 & 0.0400 & 0.0300 & 0.0250 & 0.0200 \\
\midrule 
Local loss & 0.17276 & \textbf{0.10659} & 0.05207 & \textbf{0.03384} & 0.01751 \\
\textbf{+ Global loss} & \textbf{0.17232} & \textbf{0.10659} & \textbf{0.05196} & \textbf{0.03384} & \textbf{0.01744} \\
\midrule
\midrule
Error Rate & 0.0150 & 0.0100 & 0.0075 & 0.0050 & 0.0025 \\
\midrule
Local loss  & 0.00808 & 0.00259 & \textbf{0.00097} & 0.00039 & 0.00007        \\
\textbf{+ Global loss} & \textbf{0.00802} & \textbf{0.00254} & 0.00103 & \textbf{0.00035} & \textbf{0.00005}  \\

\bottomrule
\end{tabular}%
}
% \vspace{-20pt}
\end{table}

\begin{table}[t]
\centering

\renewcommand*{\arraystretch}{1}
\setlength{\tabcolsep}{6pt}

\caption{Comparison of logicla error rates under different model size.}
% \vspace{-10pt}
\label{tab:model_size}
\resizebox{\columnwidth}{!}{%
\begin{tabular}{lcccccc}
\toprule
Error Rate & 0.0200 & 0.0150 & 0.0100 & 0.0075 & 0.0050 \\
\midrule
503K Params & 0.01812 & 0.00860 & 0.00290 & 0.00127 & 0.00045  \\
7,911K Params & \textbf{0.01744} & \textbf{0.00802} & \textbf{0.00254} & \textbf{0.00103} & \textbf{0.00035} \\

\bottomrule
\end{tabular}%
}
\end{table}

\section{Conclusion}

In conclusion, our study presents a robust QEC decoder for rotated surface codes using machine learning and transformer models. It outperforms existing benchmarks across code distances and facilitates quick transfer learning. Key factors for this success include global decoding and large Transformer models. This work advances ML-based Transformer decoders, enhancing accuracy and speed in error correction of the surface codes and beyond.

{\small
\balance
\bibliographystyle{ACM-Reference-Format}
\bibliography{ref, mypaper}
}

\end{document}